\documentclass[pdflatex,sn-mathphys]{sn-jnl}
\usepackage{subcaption}
\usepackage{graphicx}

\jyear{2021}%

\theoremstyle{thmstyleone}%
\theoremstyle{thmstyletwo}%

\theoremstyle{thmstylethree}%

\begin{document}

\title[Article Title]{Simulation of CO2 Storage using a Parameterization Method for Essential Trapping Physics: FluidFlower Benchmark Study}


\author*[1,2,3]{\fnm{Yuhang} \sur{Wang}}\email{wangyuhang17@cug.edu.cn}

\author*[2,3]{\fnm{Ziliang} \sur{Zhang}}\email{z.zhang-15@tudelft.nl}

\author*[2]{\fnm{Cornelis} \sur{Vuik}}\email{c.vuik@tudelft.nl}

\author*[3]{\fnm{Hadi} \sur{Hajibeygi}}\email{h.hajibeygi@tudelft.nl}

\affil[1]{\orgdiv{School of Environmental Studies}, \orgname{China University of Geosciences}, \city{Wuhan}, \postcode{430078}, \country{China}}



\affil[2]{\orgdiv{Faculty of Electrical Engineering, Mathematics and Computer Science, Department of Applied Mathematics}, \orgname{Delft University of Technology}, \city{Delft}, \postcode{2628XE}, \country{The Netherlands}}

\affil[3]{\orgdiv{Faculty of Civil Engineering and Geosciences, Department of Geoscience and Engineering}, \orgname{Delft University of Technology}, \city{Delft}, \postcode{2628CV}, \country{The Netherlands}}


\abstract{An efficient compositional framework is developed for simulation of CO2 storage in saline aquifers during a full-cycle injection, migration and post-migration processes.
Essential trapping mechanisms, including structural, dissolution, and residual trapping, which operate at different time scales are accurately captured in the presented unified framework.
In particular, a parameterization method is proposed to efficiently describe the relevant physical processes.
The proposed framework is validated by comparing the dynamics of gravity-induced convective transport with that reported in the literature.
Results show good agreement for both the characteristics of descending fingers and the associated dissolution rate. 
The developed simulator is then applied to study the FluidFlower benchmark model. An experimental setup with heterogeneous geological layers is discretized into a two-dimensional computational domain where numerical simulation is performed.
Impacts of hysteresis and the diffusion of CO2 in liquid phase on the migration and trapping of CO2 plume are investigated.
Inclusion of the hysteresis effect does not affect plume migration in this benchmark model, whereas diffusion plays an important role in promoting convective mixing.
This work casts a promising approach to predict the migration of the CO2 plume, and to assess the amount of trapping from different mechanisms for long-term CO2 storage.}

\keywords{Geological CO2 storage, Porous media, Compositional simulation, Convective mixing}

\maketitle

\section{Introduction}\label{sec1}

Carbon dioxide capture and storage (CCS) is one of promising measures to reduce greenhouse gas emission, and mitigate the climate change. This technique usually consists of subsequent processes, including collection of CO2 emission from large point source like power plants, transportation of the collected gas emission, and storing it in geological formations \cite{metz2005ipcc}. Of potential geological targets, deep saline aquifers are of great importance because they are prevalent in sedimentary basins and have promising storage capacity. However, the spatial heterogeneity in geological formations, as well as high contrast between fluids when storing CO2 in saline aquifers, pose challenges on understanding and simulating such unstable processes \cite{wang2019extension,boon2022anisotropic,ershadnia2020co2,jackson2020small,zheng2021hybrid,kou2022method,shao2022modelling,wang2022analysis}. 

The well classified trapping mechanisms, by which injected CO2 can be sequestrated in pore spaces, include structural and stratigraphic trapping \cite{emami2015convective,andersen2017simplified}, residual trapping \cite{hunt1988nonaqueous,juanes2006impact,herring2013effect}, dissolution trapping \cite{duan2003improved,spycher2003CO2,spycher2005CO2}, and mineral trapping \cite{dai2020reactive}. Structural and stratigraphic trapping is the part of CO2 trapped in some topological features or stratigraphic traps formed by change of rock type, as CO2 migrates in the reservoir. Residual trapping, also known as capillarity trapping, refers to the CO2 that is left behind and loses its spatial continuity when water reinvades into pore space occupied by CO2. Such trapping occurs mostly in areas near wellbore and plumes spreading under caprock after injection stops. Dissolution trapping occurs when CO2 and brine are in contact with each other during injection and migration. The amount of CO2 can be dissolved into brine relies on pressure, temperature, and salinity of brine, and their effects on CO2 dissolubility have been extensively studied in literature. Mineral trapping occurs the process during which dissolved CO2 is precipitated as carbonate minerals and trapped in this more permanent form. However, because of the slow rate of geochemical reaction, this trapping mechanism is usually considered when the involved process spreads over large time scales. 

The relative importance of different trapping mechanisms varies temporally \cite{ide2007storage}. As governed by advective transport during injection period, structural and stratigraphic trapping dominates the contribution to stored CO2 at early stage. Residual trapping comes into effect when injection stops, since most transitions of the displacement process, i.e., from drainage to imbibition, take place at this point in time. The above two trapping mechanisms operate at a time scale close to injection period, while the dissolution trapping lasts longer \cite{wang2022analysis}. 
Dissolution occurs in regions swept by CO2 during injection, and continues to play a crucial role due to the gravity-induced convective transport in post-injection period \cite{ranganathan2012numerical,szulczewski2013carbon,emami2015two,de2017dissolution,wen2021convective}.
Obviously, mineral trapping operates at a even larger time scale than dissolution because of the slow reaction kinetics \cite{gunter2000aquifer,juanes2006impact}. 
In order to evaluate the interplay of those trapping mechanisms, it is of necessity to develop an efficient and robust numerical framework. 
One of the challenging tasks is to capture the hysteresis effect. 
Especially in the CO2-brine system, the hysteretic behaviour of relative permeability and capillary pressure has already been revealed by experimental studies \cite{oak1990three, akbarabadi2013relative, ruprecht2014hysteretic, pini2017capillary}. 
Despite of physical observations, the residual gas is usually accounted for using a fixed value in numerical simulations.
In this study, the history of saturation is considered when determining residual gas saturation, and its value is calculated based on the turning point at which the transition happens \cite{land1968calculation,steffy1997influence}.

Efforts have been made to investigate CO2 storage using numerical models. In most case, simulations are conducted at field-scale and coarse grids are often employed \cite{class2009benchmark,nordbotten2012uncertainties}. In consequence, the numerical error, as well as the spatial heterogeneity may overwhelm  the relevant physical processes, and it is impractical to compare the simulation results with any physical observations.
The FluidFlower benchmark study provides an excellent opportunity to examine the predictive capability of the Darcy-scale model, and to study the interactions of underlying physics, some of which may be neglected at field-scale (e.g., CO2 diffusion in liquid phase), in a comprehensive manner. 
The experiment is designed to mimic CO2 storage in a layered system using a meter-scale rig. Physical parameters of different sand types are measured which will be used in the simulation. 
The uncertainty arsing from either spatial heterogeneity or physical properties are well-regulated, allowing for a direct comparison between numerical predictions and experimental observations.


This work is structured as follows. we first present the governing equations describing the CO2-brine system, thermodynamic equilibrium calculation, and the solution strategy. Following that we show how physical models are parameterized to improve the computational efficiency.
The proposed compositional framework is validated by investigating the gravity-induced convective transport. Next, we present simulation results of FluidFlower benchmark model, and discuss the impacts of key physical processes on plume migration and trapping of CO2. Key findings are summarized at the end.

\section{Methodology} 

\subsection{Compositional formulation} 

The governing equation describing a multi-component, multi-phase flow system is given by:
\begin{equation} \label{Mass conservation}
\frac{\partial}{\partial t}\left(\phi\sum_{\alpha=1}^{n_{ph}}{x_{c,\alpha}\rho_\alpha S_\alpha}\right)
+\nabla\cdot\sum_{\alpha=1}^{n_{ph}}\left({x_{c,\alpha}\rho_\alpha u_\alpha} + S_\alpha \rho_\alpha J_\alpha\right) 
= \sum_{\alpha=1}^{n_{ph}}{x_{c,\alpha}\rho_\alpha q_\alpha},
\end{equation}
where subscript $\alpha$ represents liquid (brine) or gas (supercritical CO2) phase, and $c$ denotes the component,i.e., brine or CO2 in this case. $\phi$ is rock porosity; $\rho_{\alpha}$, $S_{\alpha}$, and $q_{\alpha}$ are the molar (mass) density, saturation, and flow rate of phase $\alpha$, respectively. Note that $x_{c,\alpha}$ is the molar (mass) fraction of component $c$ in phase $\alpha$. $u_\alpha$ is the phase velocity and described by the classic extension of Darcy's law:
\begin{equation}
u_\alpha=\frac{kk_{r,\alpha}}{\mu_\alpha}\nabla\left(p_\alpha-{\rho}_\alpha g h\right), 
\end{equation}
where $k$ is the rock permeability; $k_{r\alpha}$ and $\mu_\alpha$ are phase relative permeability and viscosity, respectively. $J_\alpha$ is the diffusive flux of phase $\alpha$. According to Fick's law, $J_\alpha$ is proportional to the gradient of molar (mass) fraction given by
\begin{equation}
J_\alpha=-\phi D_{c,\alpha}\nabla x_{c,\alpha},
\end{equation}
where $D_{c,\alpha}$ is the diffusion coefficient of component $c$ in phase $\alpha$. 


\subsection{Overall-compositional formulation} 

In our two-component two-phase system, the overall composition variable set is chosen to solve the system of nonlinear equations, using phase pressure and overall molar fraction as primary variables \cite{voskov2012comparison}. The overall mole friction of component $c$, $z_c$, is defined as:
\begin{equation}
z_c=\sum_{\alpha=1}^{n_{ph}}{\nu_\alpha x}_{c,\alpha},\ \ \forall c\in{1,\ ..,n_c}
\end{equation}
where $\nu_\alpha=\frac{S_\alpha\rho_\alpha}{\sum_{\alpha=1}^{n_{ph}}{S_\alpha\rho_\alpha}}$ represents the molar (mass) fraction of phase $\alpha$. With the defined overall molar fraction term, the mass conservation equation for each component can be rewritten as:
\begin{equation}
\frac{\partial}{\partial t}\left(\phi\rho_Tz_c\right)
+\nabla\cdot\sum_{\alpha=1}^{n_{ph}}\left({x_{c,\alpha}\rho_\alpha u_\alpha} + \rho_T J_\alpha\right) 
= \sum_{\alpha=1}^{n_{ph}}{x_{c,\alpha}\rho_\alpha q_\alpha},
\end{equation}
here $\rho_T$ denotes the total density, and it is a function of phase density and saturation.

\subsection{Analytical flash calculation} 

When multiple phases are present, it is usually assumed that all phases reach  thermodynamic equilibrium instantaneously, in order to decouple phase behaviour calculation from flow \cite{cusini2018algebraic, lyu2021numerical}. In this work, we consider a black-oil type fluid model which allows the lighter component to exit in both phases. The equilibrium ratio, also known as the $k$-value, determines how one component is split into different phases:
\begin{equation}
k_c=\frac{x_{c,g}}{x_{c,l}}.
\end{equation}
The phase constraints based on molar fraction terms read:
\begin{equation}
\sum_{c=1}^{n_c}x_{c,\alpha}=1,\   \ \sum_{c=1}^{n_{ph}}\nu_{\alpha}=1.
\end{equation}
Based on the definition of the overall molar fraction term, phase distribution parameters can be derived analytically as:
\begin{equation}
\left[\begin{matrix}
x_{co_2,g} & x_{b,g} & V \\ \\ 
x_{co_2,l} & x_{b,l} & L \\
\end{matrix}\right]
=
\left[\begin{matrix}
1 & 0 & \frac{z_{co_2} - \frac{1}{k_{co_2}}}{1-\frac{1}{k_{co_2}}} \\ \\
\frac{1}{k_{co_2}}&1-\frac{1}{k_{co_2}} & \frac{1- z_{co_2}}{1-\frac{1}{k_{co_2}}} \\
\end{matrix}\right].
\end{equation}

\subsection{Solution strategy} 

The nonlinear system of equations is solved using finite volume discretization in space and implicit discretization in time. The nonlinear equation for component $c$ in cell $i$ in residual form can be linearized with the Newton-Raphson method as:
\begin{equation}
r_{c,i}^{v+1} \approx 
r_{c,i}^v + \left(\frac{\partial r_{c,i}}{\partial p_l}\right)^v \delta p_l^{v+1}+ \left(\frac{\partial r_{c,i}}{\partial z_{co_2}}\right)^v \delta z_{co_2}^{v+1}=0,
\end{equation}
here $v$ and $v+1$ are the current and next iteration step, respectively. Then the linearized equations is solved in an iterative manner:
\begin{equation}
J^v\delta x^{v+1}=-r^v,
\end{equation}
where $J^v$ and $r^v$ are Jacobian and Residual matrix, and $\delta x^{v+1}$ is the primary variable update. The same function in matrix multiplication form is given  below:
\begin{equation}
\left(\begin{matrix}\frac{\partial r_{co_2}}{\partial p_l}&\frac{\partial r_{co_2}}{\partial z_{co_2}}\\ \\ \frac{\partial r_b}{\partial p_l}&\frac{\partial r_b}{\partial z_{co_2}}\\\end{matrix}\right)\left(\begin{matrix}\delta p_l\\ \\ \delta z_{co_2}\\\end{matrix}\right)=-\left(\begin{matrix}r_{co_2}\\ \\ r_b\\\end{matrix}\right).
\end{equation}
The process is repeated until it reaches non-linear convergence, i.e., the infinite norm of the residual and variable update are less than a certain tolerance. In addition, an adaptive time stepping strategy is employed to reduce the number of iterations needed to reach convergence. In the implementation, the time step size is dynamically changed between user defined maximum and minimum time step size, based on the number of iterations needed to converge.

\section{Parameterized physical models} 


\subsection{Dissolution} 
The solubility of CO2 in brine varies with pressure, temperature and water salinity \cite{wang2022analysis}. In this work, the volume of CO2 that can be dissolved into a unit volume of brine is quantified by the solution CO2-brine ratio, or $R_s$, at given temperature and salinity. As a result, $R_s$ depends on pressure only. The corresponding parameterization space can be readily extended spreading across different temperature and salinity settings nevertheless. 

We follow a thermodynamic model that equates chemical potential to predict the dissolution of CO2 \cite{spycher2003CO2,hassanzadeh2008predicting}. The solution CO2-brine ratios are calculated using the mole fraction of CO2 in liquid phase obtained from the CO2 molality in brine as:
\begin{equation}
R_s=\frac{\rho_b^{\text{STC}}x_{CO2,l}}{\rho_{CO2}^{\text{STC}}(1-x_{CO2,l})}.
\end{equation}
Here $\text{STC}$ is short for standard condition. The relationship between calculated CO2 molality and pressure is verified by comparing against the data reported in literature. CO2 molality is then converted to the solution ratio, which is stored in a lookup table in the offline stage. The $R_s$ value can be directly read from the table with known pressure.

It is worth mentioning that the lookup table only applies to cells in two-phase state, i.e., CO2 exists in both gas and liquid phases. Therefore, a stability test needs to be preformed to check the number of phases in each cell after updating the primary variables at each iteration. Assuming brine presents in liquid phase only, the $k$-values for CO2 and brine are given by:
\begin{equation}
k_{co_2}=\frac{\rho_{co_2}^{STC}R_s+\rho_b^{STC}}{\rho_{co_2}^{STC}R_s},\ \ k_b=0.
\end{equation}
If a cell is in the state of liquid phase solely, the amount of CO2 dissolved in brine is not enough to reach the dissolution limit \cite{hajibeygi2014compositional}. In other words, the solution CO2-brine ratio needs to be modified with the overall molar fraction of CO2 given by:
\begin{equation}
R_s=\frac{\rho_b^{STC}z_{CO2}}{\rho_{CO2}^{STC}(1-z_{CO2})}.
\end{equation}

\subsection{Hysteresis} 

\begin{figure}
  \centering
  \includegraphics[width=0.8\linewidth]{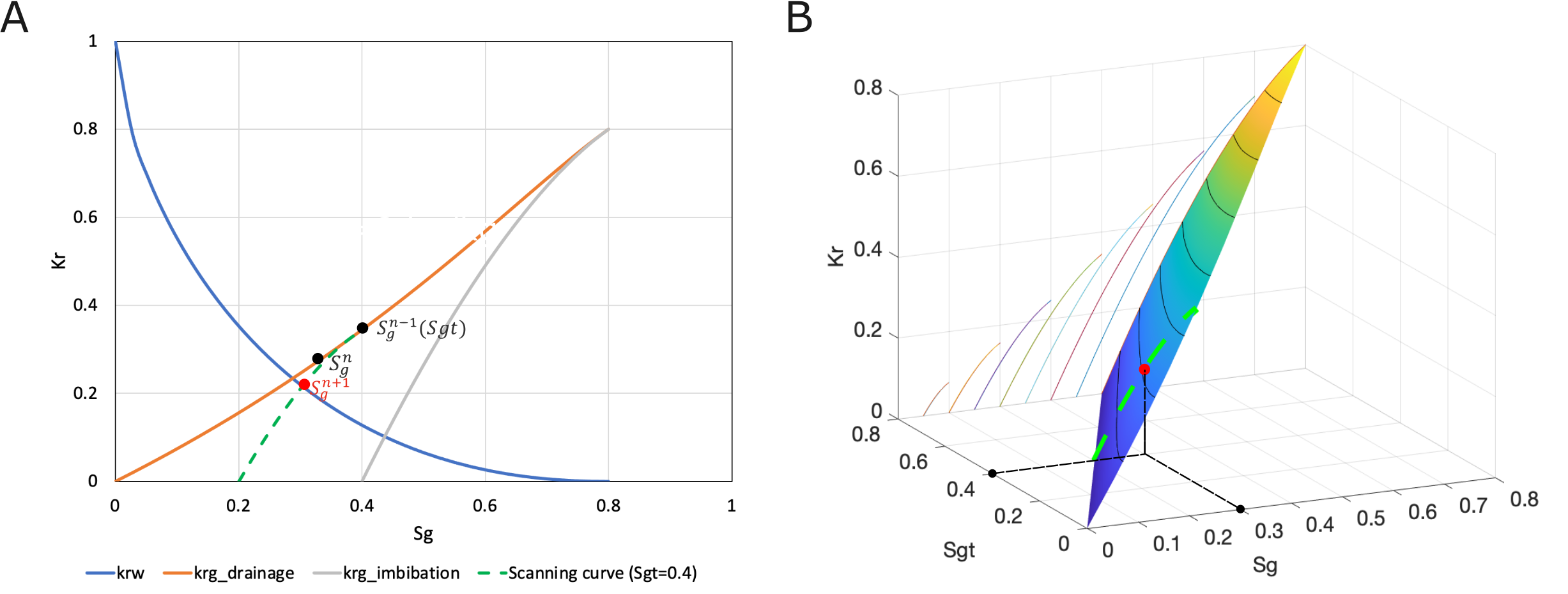}
  \caption{Illustration of hysteretic relative permeability models. (A) Determination process based on scanning curves. (B) Determination process based on the proposed scanning curve surface.}
\label{fig:Scanning curve and surface}
\end{figure}

The hysteresis effect mainly refers to the behavior that different constitutive relations, i.e., relative permeability and capillary pressure curves, are followed depending on the type of on-going process, i.e., drainage or imbibition. The point where transition happens is called turning point, $S_{gt}$.
In previous work, a series of scanning curves starting from different turning points, are constructed to ensure the transition is continuous \cite{killough1976reservoir, wang2022analysis}. The scanning curve for each cell is constructed individually after the flow process is determined. The determination of the process is based on comparing gas saturation from the previous two time steps, $n$ and $n-1$. For example, if $S_g^n<S_g^{n-1}$ on primary drainage curve, it indicates the process has already been transitioned to the scanning curve at  $S_g^{n-1}$, as shown by \autoref{fig:Scanning curve and surface}A.

As mentioned earlier, the scanning curves were constructed cell by cell, which may lead to repetitive construction of scanning curves.
To improve the computational efficiency, a parameterized space, or scanning curve surface, across different turning points and saturation values, are constructed (in the offline stage). In the newly proposed workflow, the type of process for each cell is determined in a vectorized manner without looping each cell. The gas relative permeability or capillary pressure are then read directly from the surface, knowing the pair of gas saturation and turning point, as shown in \autoref{fig:Scanning curve and surface}B.
\begin{figure}
  \centering
  \includegraphics[width=0.8\linewidth]{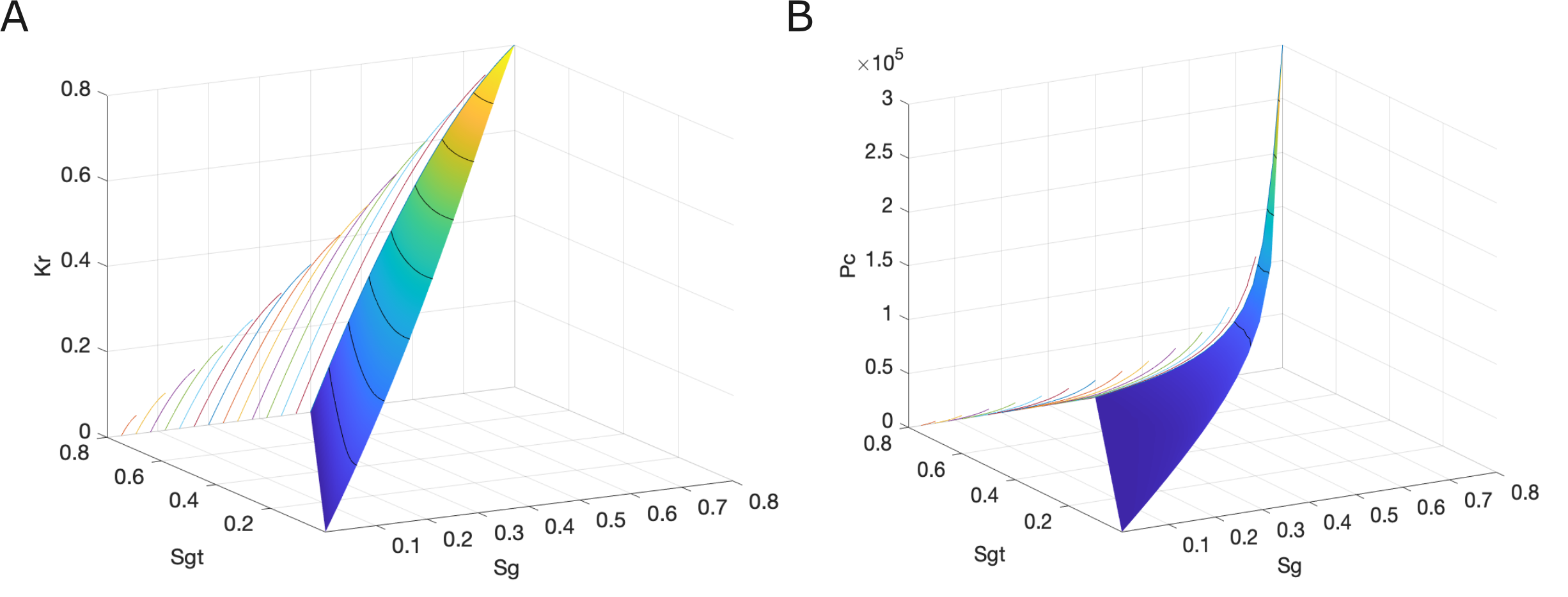}
  \caption{Scanning curve surfaces to model the hysteretic behaviors of constitutive relations. (A) Relative permeability. (B) Capillary pressure.}
\label{fig:Scanning curve surfaces}
\end{figure}

\subsection{Capillarity} 

Capillary pressure function below relates the pressure difference between different phases as:
\begin{equation} \label{Pc}
p_\alpha-p_\beta={p_c}_{\alpha,\beta},\ \ \forall\alpha\neq\beta\in{1,\ ..,n_{ph}}.   
\end{equation}
Capillary pressure plays an important role in plume migration and CO2 trapping.
CO2 can be collected under structural and stratigraphic traps when the buoyancy forces cannot overcome the capillary forces in the narrow pore throat of caprock. This ensures CO2 will not enter the overlying pore space. In addition, a portion of CO2 is left as discontinuous ganglia and becomes immobilized because of capillary forces. This is referred to as residual trapping, which is affected by pore structure and wettability of the rock. As shown by \autoref{fig:Scanning curve surfaces}B, capillary pressure and its hysteresis behaviour are also captured by the parameterized space.

\section{Validation} 
\begin{figure}
  \centering
  \includegraphics[width=0.8\linewidth]{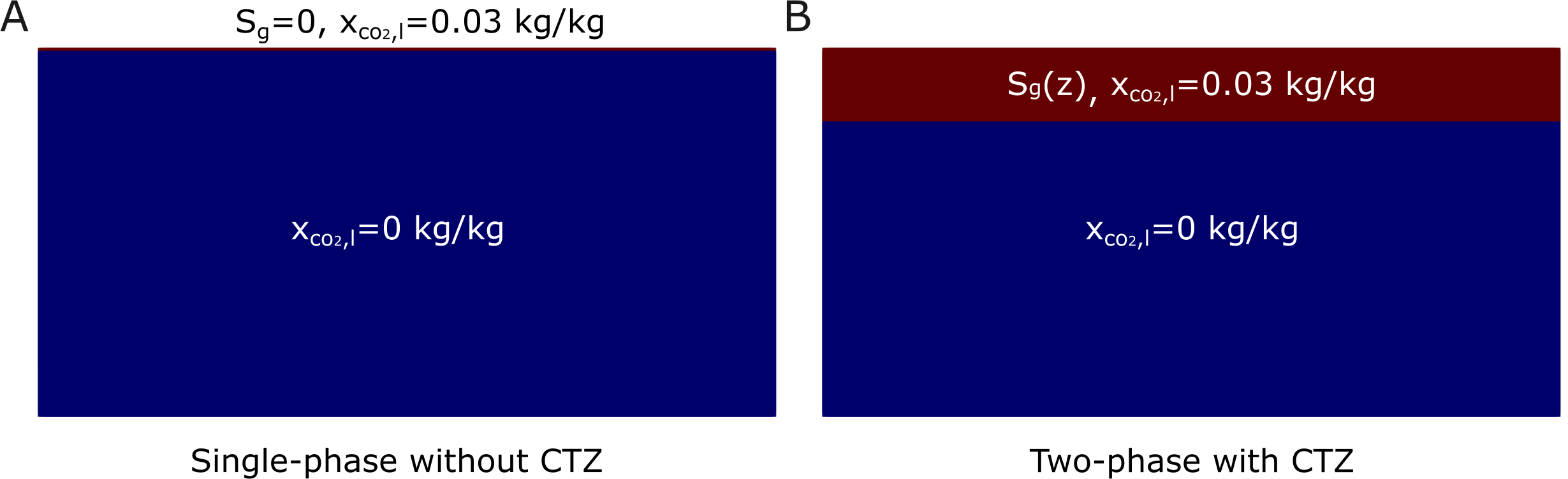}
  \caption{Initial distribution of pure brine (blue) and single-phase brine dissolved with CO2 (red) in case A, and two-phase region with free gas (red) in case B.}
\label{fig:Case A & B}
\end{figure}

In this section, we validate the proposed compositional framework by investigating a 2D synthetic model. The behaviour of gravity-induced convective transport, and the associated dissolution rate in the absence and presence of capillarity transition zone (CTZ) \cite{elenius2014convective, elenius2015interactions}, are studied in details. 
Two simulation cases we employ are a) single-phase without CTZ, and b) two-phase with CTZ, which represent negligible and stagnant CTZ scenarios, respectively (shown in \autoref{fig:Case A & B}). 
In the case of single-phase flow (case A), cells of the top boundary have CO2 concentration fixed at solubility limit by specifying a large pore volume. 
On the other hand, cells in the top 10 m of case B are saturated with gaseous CO2. 
In both cases, no-flow condition is imposed on all boundaries, and the bottom part of domain is fully saturated with single-phase brine at initial stage. The parameters of reservoir and fluid properties are summarized in \cite{elenius2015interactions}.

\begin{figure}
  \centering
  \includegraphics[width=0.8\linewidth]{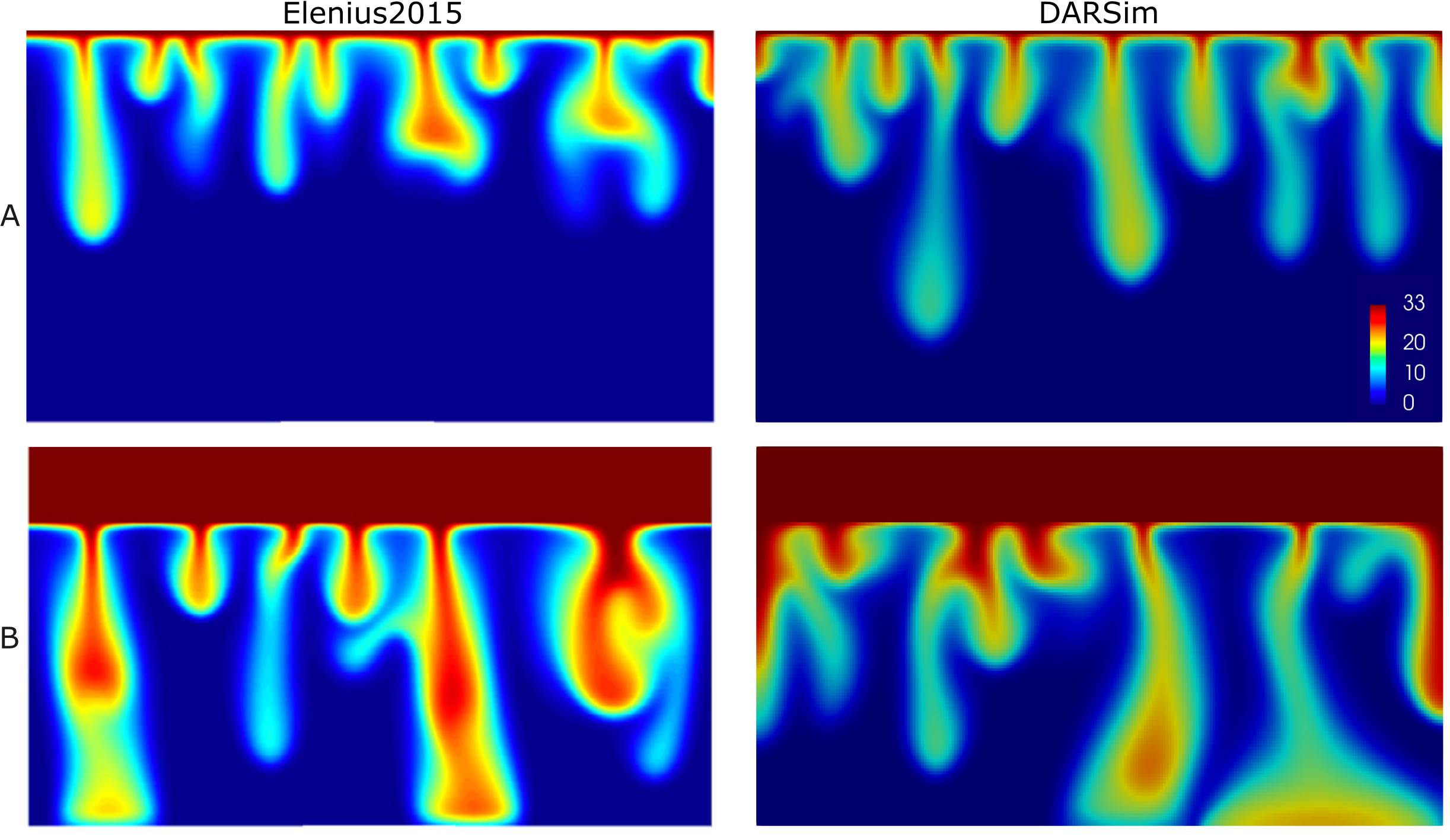}
  \caption{CO2 concentration profiles of case A and B at 200 years from literature and DARSim.}
\label{fig:Case A & B_results}
\end{figure}

\autoref{fig:Case A & B_results} presents CO2 concentration profiles after 200 years in case A and B. It clearly reveals that CTZ accelerates the propagation of fingers and causes more pronounced convective transport, as shown by the fact that fingers in case B propagate much further with larger concentration at 200 years. The above findings are consistent with results in literature. On the other hand, results obtained by the proposed compositional framework show good agreement with the maps reported in literature, in terms of the characteristics of fingers, such as width and shape. The slightly further development of convective fingers in our simulation results can be attributed to the randomness in perturbation \cite{elenius2015interactions}.

\begin{figure}
  \centering
  \includegraphics[width=0.9\linewidth]{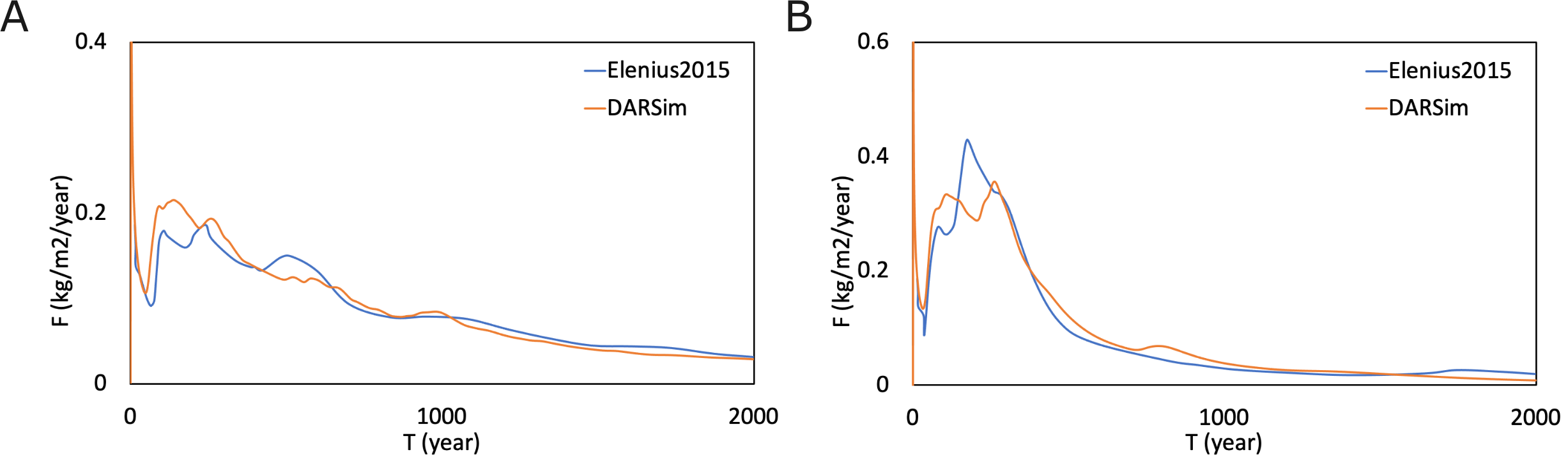}
  \caption{Dissolution rate of case A and B over 2000 years.}
\label{fig:Case A & B_F}
\end{figure}

To quantify the mass transfer rate of CO2 from regions above to single-phase brine at bottom, dissolution rate $F$ is introduced in literature, which is given by:
\begin{equation}
F = h \phi \frac{\partial \overline{c}}{\partial t}.
\end{equation}
Here $h$ and $\phi$ denote the height and porosity of single-phase brine region, respectively. $\overline{c}$ is the mean CO2 concentration of the same region. As given by \autoref{fig:Case A & B_F}, dissolution rates of both cases follow well with the results obtained from the previous study. As shown, the dissolution rate decreases at the early start because the process is dominated by diffusion. After the so-called onset time, $F$ starts to increase due to the fact that perturbations grow into descending fingers which enhances the mass flux. The dissolution rate in both two cases fluctuates around constant values until fingers reach the bottom of the aquifer. At the end, $F$ gradually decreases as fingers merge with each other and the driving force for downward transport decreases \cite{tsinober2022role}.

\begin{figure}
  \centering
  \includegraphics[width=0.9\linewidth]{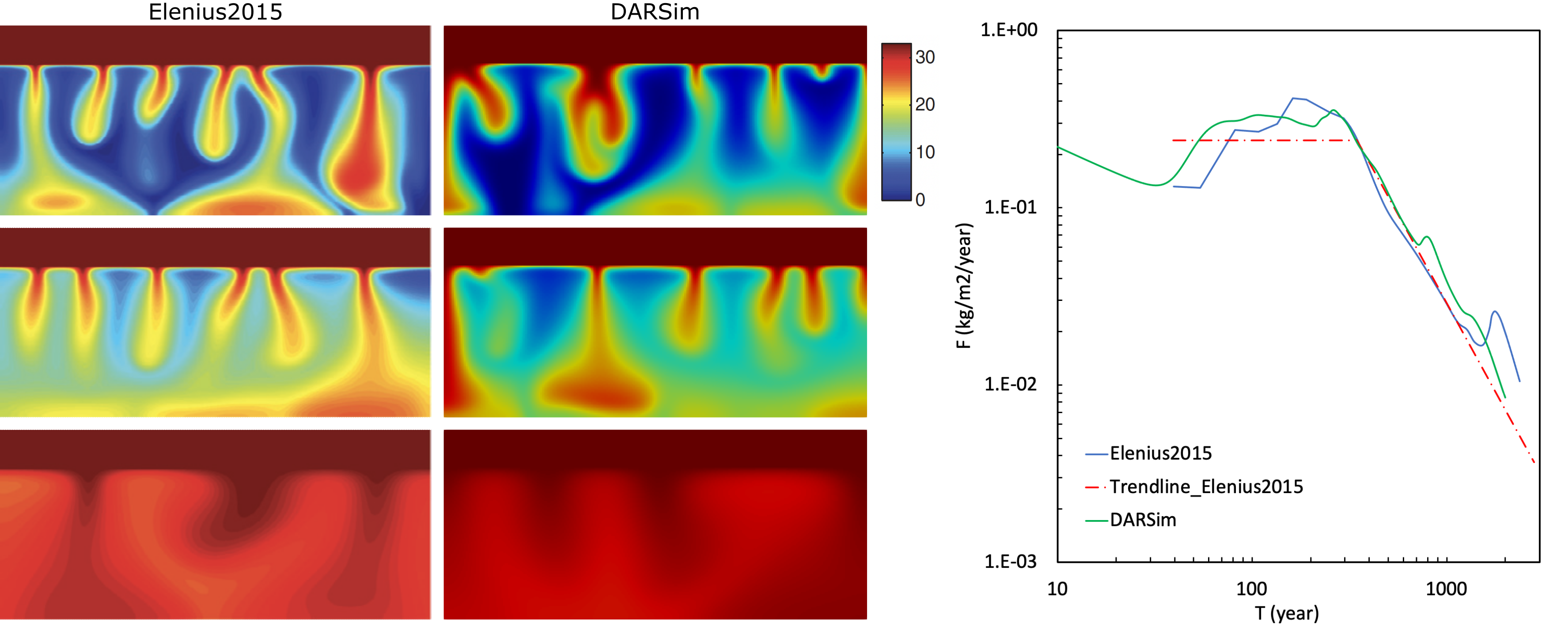}
  \caption{CO2 concentration profiles of case B after 250 (top), 350 (middle), and 2000 (bottom) years, and its associated dissolution rate.}
\label{fig:Case B_results and F}
\end{figure}

\autoref{fig:Case B_results and F} shows the distribution of CO2 concentration after 250, 350 and 2000 years of stagnate CTZ case agree well with reported data in literature. \autoref{fig:Case B_results and F} also plots the associated dissolution rate in logarithm scale against the analytical solution. Simulation results capture the convective transport and fit well with the prediction of analytical solution. Specifically, we found a similar $t_{peel}=350$ years, before which dissolution rate stabilizes around a constant value. After fingers start propagate along aquifer bottom, dissolution rate becomes proportional to $1/t^2$ \cite{elenius2015interactions}.

\section{FluidFlower Benchmark Study}
The validated compositional framework is applied to the FluidFlower benchmark study, in which numerical predictions are compared with experimental observations. Impacts of key physical effects on CO2 storage are examined.

\subsection{Model description}
\begin{figure}
  \centering
  \includegraphics[width=0.8\linewidth]{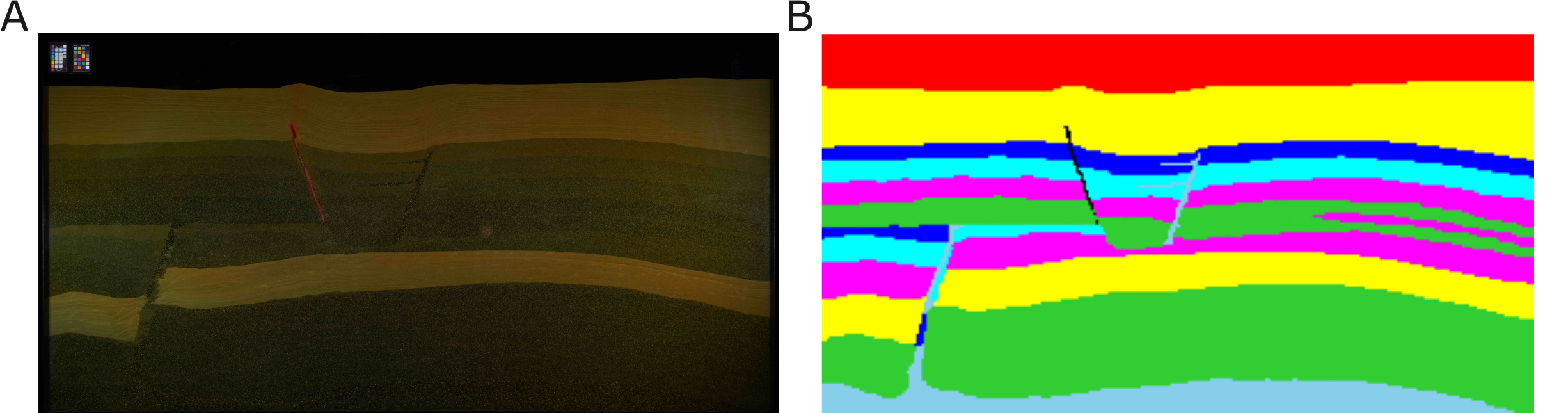}
  \caption{Geometry of FluidFlower model. (A) Original geometry. (B) Discretized geometry.}
\label{fig:FluidFlower geometry}
\end{figure}

FluidFlower setup is designed to accommodate multiphase 2D flow experiments and to reproduce key processes happening during CO2 storage. Thus, a direct visualization of flow dynamics, which is missing from numerical studies, can be delivered. The geometry of geological layers, conceived based on typical North Sea reservoirs, is formed by pouring unconsolidated sand with varying types separately into the water-filled experiment rig. After careful sedimentation operation, the geometry was flushed with different aqueous fluids prior to CO2 injection. Together with the free water table above used to stabilize pressure, the length and height of the setup are 2.86 m and 1.53 m respectively, while the width varies between 0.019 m and 0.028 m from two sides to the center. Simulations are performed in a 2D domain discretized by 286$\times$1$\times$153 cells, assuming the width is fixed at 0.019 m. \autoref{fig:FluidFlower geometry} shows that the geometry used in the experiment is well represented by the discretized model. 

\begin{figure}
  \centering
  \includegraphics[width=0.8\linewidth]{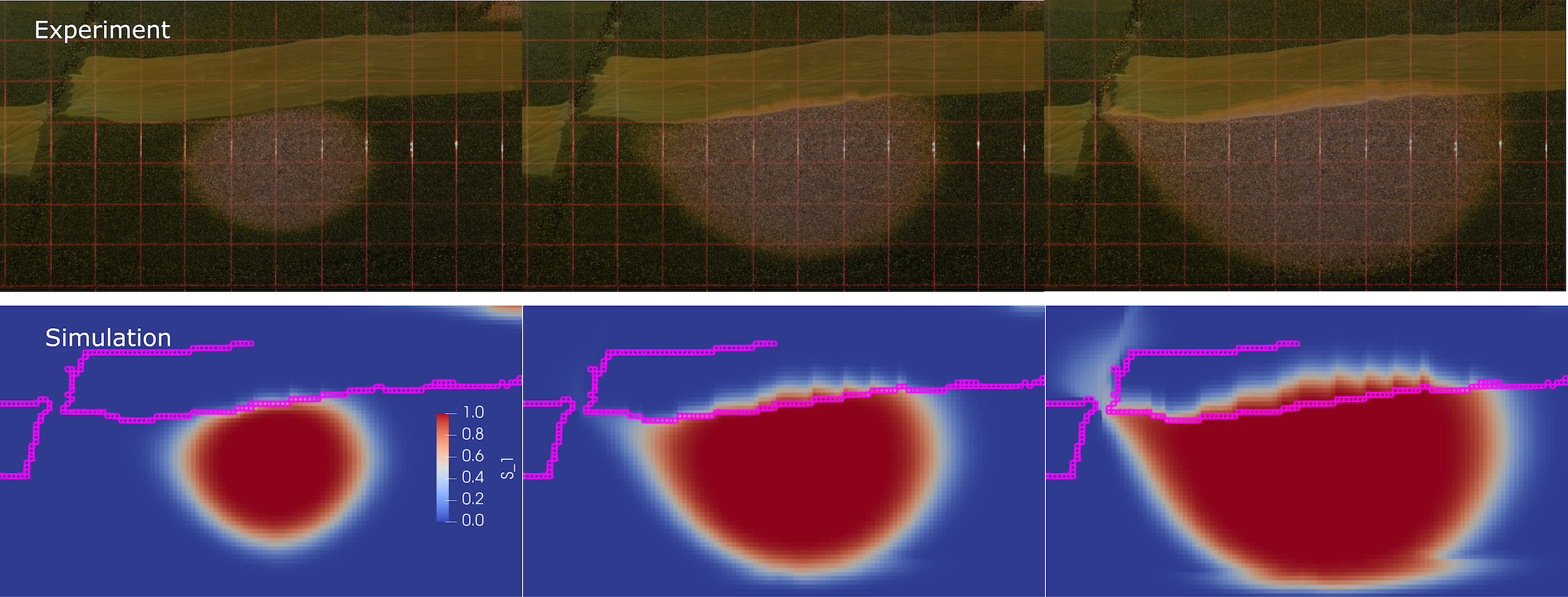}
  \caption{Snapshots of tracer test through lower injection port in experiments and simulation (Sealing layer boundary cells are highlighted in simulation results for comparison).}
\label{fig:Tracer test_2}
\end{figure}

Permeability values of different sand layers are tuned such that the simulation results match those obtained from the tracer test. During the tracer test, aqueous phases with different colors are injected through two injection ports with constant volumetric flow rate and square pulses pattern. \autoref{fig:Tracer test_2} shows that the evolution of tracer front agrees well between experimental observations and simulation results with tuned permeability field.




\begin{figure}
  \centering
  \includegraphics[width=0.8\linewidth]{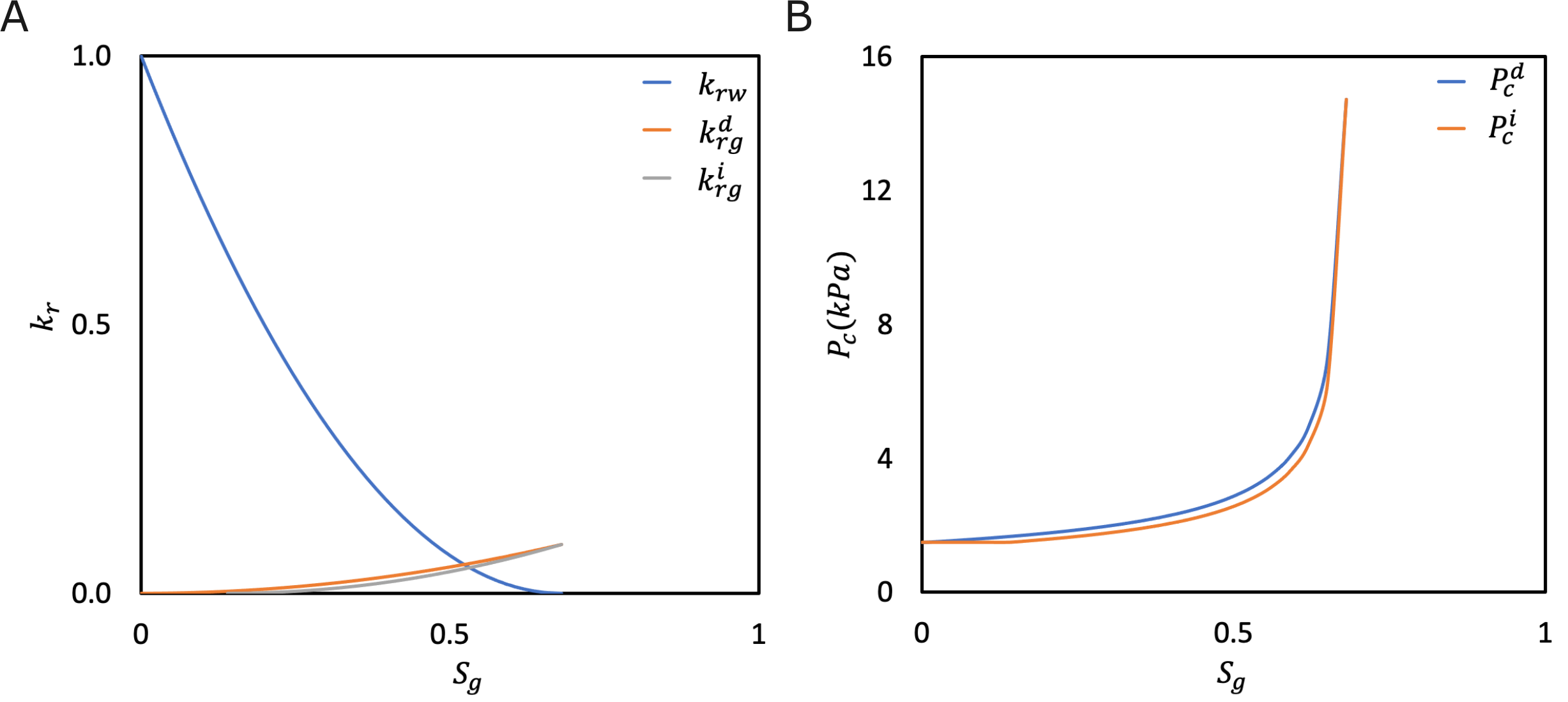}
  \caption{Relative permeability (A) and capillary pressure model (B) of sand ESF.}
\label{fig:kr and Pc}
\end{figure}

The relative permeability and capillary pressure curves of different sand types are constructed using the Brooks-Corey model, in which the effective saturation can be computed based on residual saturation values as:
\begin{equation}
S_{ge}=\frac{S_g-S_{gr}}{1-S_{gr}-S_{wr}}.
\end{equation}
Relative permeability are then calculated by
\begin{equation}
k_{rw}=k_{rw}^0\left(1-S_{ge}\right)^2,\ \ k_{rg}=k_{rg}^0 S_{ge}^2,
\end{equation}
where $k_{rw}^0$ and $k_{rg}^0$ are relative permeability end points. Similarly, capillary pressure for sand ESF, C and D are given by:
\begin{equation}
P_c=P_{ce}\left(1-S_{ge}\right)^{-0.5}.
\end{equation}
Using sand type ESF as an illustrative example, \autoref{fig:kr and Pc} shows the hysteretic behaviour of gas phase can be described by the proposed parameterization method, while the same curve is used for liquid phase in drainage and imbibition processes.

Neglecting the thermal effect, the system temperature is assumed to be constant at $20^{\circ}$C. The reservoir is initialized with atmospheric pressure. Particularly, pressure at the top boundary is fixed since it is open and is connected to a free water table. No flow occurs across other boundaries. In our study, the injection scheme is implemented following the experimental setting, during which CO2 is injected with a constant rate of 10 ml/min through two injection ports for 5 hours and 2 hours 45 mins respectively. After injection ceases, simulation continues until 5 days to predict the migration and fate of CO2 plume. Fluid properties and some physical parameters used in the simulations are summarized in \autoref{tab:Physical parameters_Johansen}.
\begin{table}[]
\begin{center}
\caption{Physical parameters and simulation scheme of FluidFlower simulation cases.}\label{tab:Physical parameters_Johansen}
\begin{tabular}{@{}llll@{}}
\toprule
Properties & Symbols & Values & Units\\
\midrule
{Reservoir temperature}                        & $T$           & 293.15    & K       \\
{Initial reservoir pressure}                   & $p_0$         & 101325    & Pa     \\
{CO2 viscosity}                                & $\mu_{\text{CO}2}$  & 1.77e-5   & Pa.s   \\
{Brine viscosity}                              & $\mu_{b}$     & 1.01e-3   & Pa.s    \\
{Injection rate}                               & $q$           & 10        & ml/min \\
{Injection time}                               & $t_{\text{inj}}$     & 5/2.75    & hours    \\
{Simulation time}                              & $t_{\text{tot}}$     & 5         & days    \\
\botrule
\end{tabular}
\end{center}
\end{table}

\subsection{Impacts of hysteresis}
We first study the impact of hysteresis by considering the hysteretic relative permeability and capillary pressure using the proposed parameterization method. The scanning curve surfaces, which are generated based on primary drainage and imbibition curves (shown in \autoref{fig:kr and Pc}), are used to model $k_r$ and $P_c$. On the other hand, in the case where hysteresis is neglected, we model such case by assuming the drainage and imbibition processes follow the same curve. This assumption leads to zero residual gas saturation as the primary drainage curve starts from the saturation value of 0.

\begin{figure}[h]
  \centering
  \includegraphics[width=0.9\linewidth]{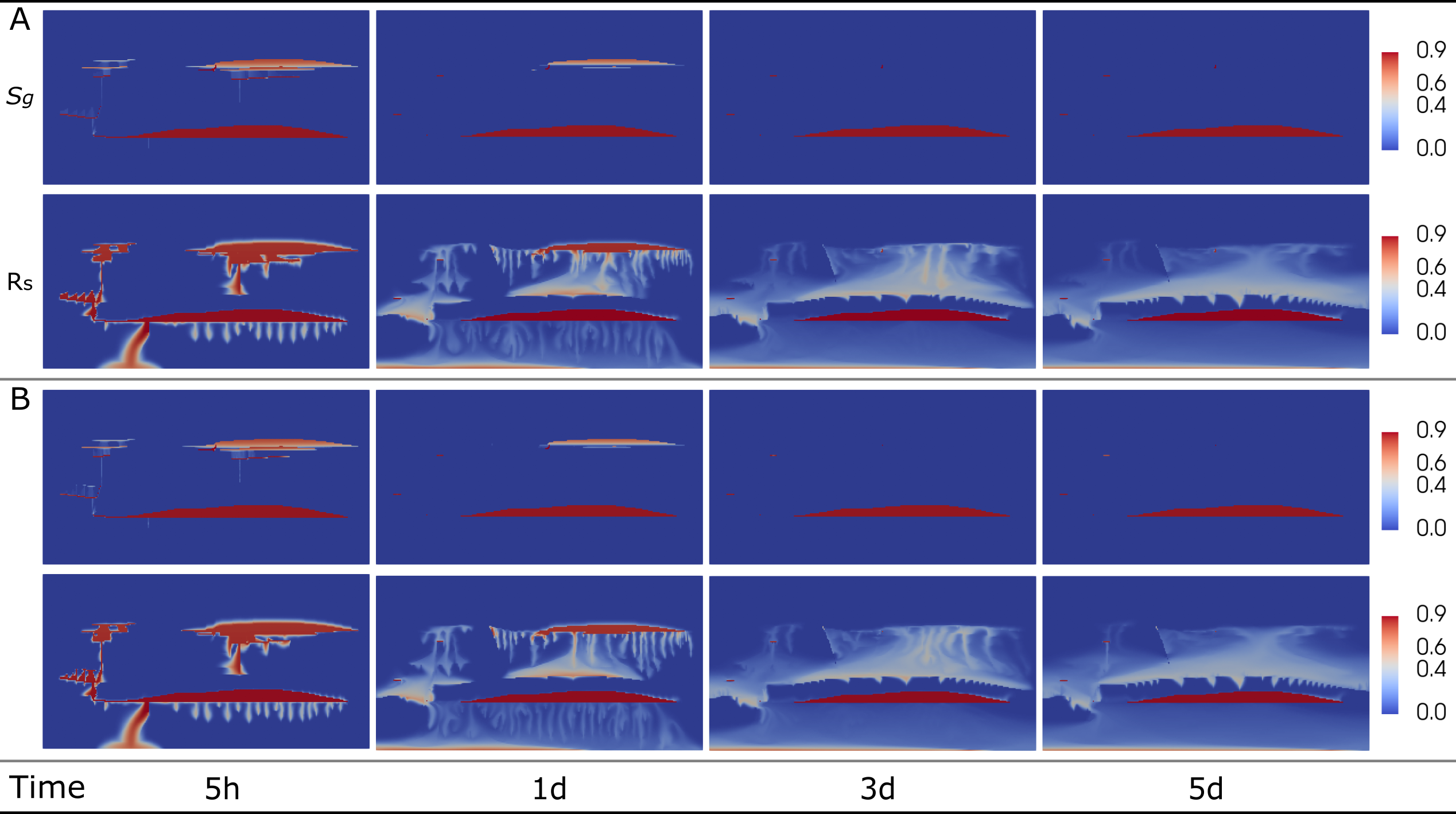}
  \caption{Gas phase saturation and solution CO2–brine ratio profiles obtained from the case (A) without and (B) with hysteresis effect.}
\label{fig:Hysteresis results}
\end{figure}

\autoref{fig:Hysteresis results} shows CO2 saturation and dissolution ratio profiles at several representative time steps in the absence and presence of hysteresis effect, respectively. Minor differences between the two sets of maps indicate that including hysteresis does not have a pronounced impact.  
This is due to the fact that the primary drainage and imbibition curve is very close to each other. Therefore, even though flow may transition from primary curve to the scanning ones, the deviation of relative permeability (or capillary pressure) value is insignificant.

\begin{figure}[h]
  \centering
  \includegraphics[width=0.8\linewidth]{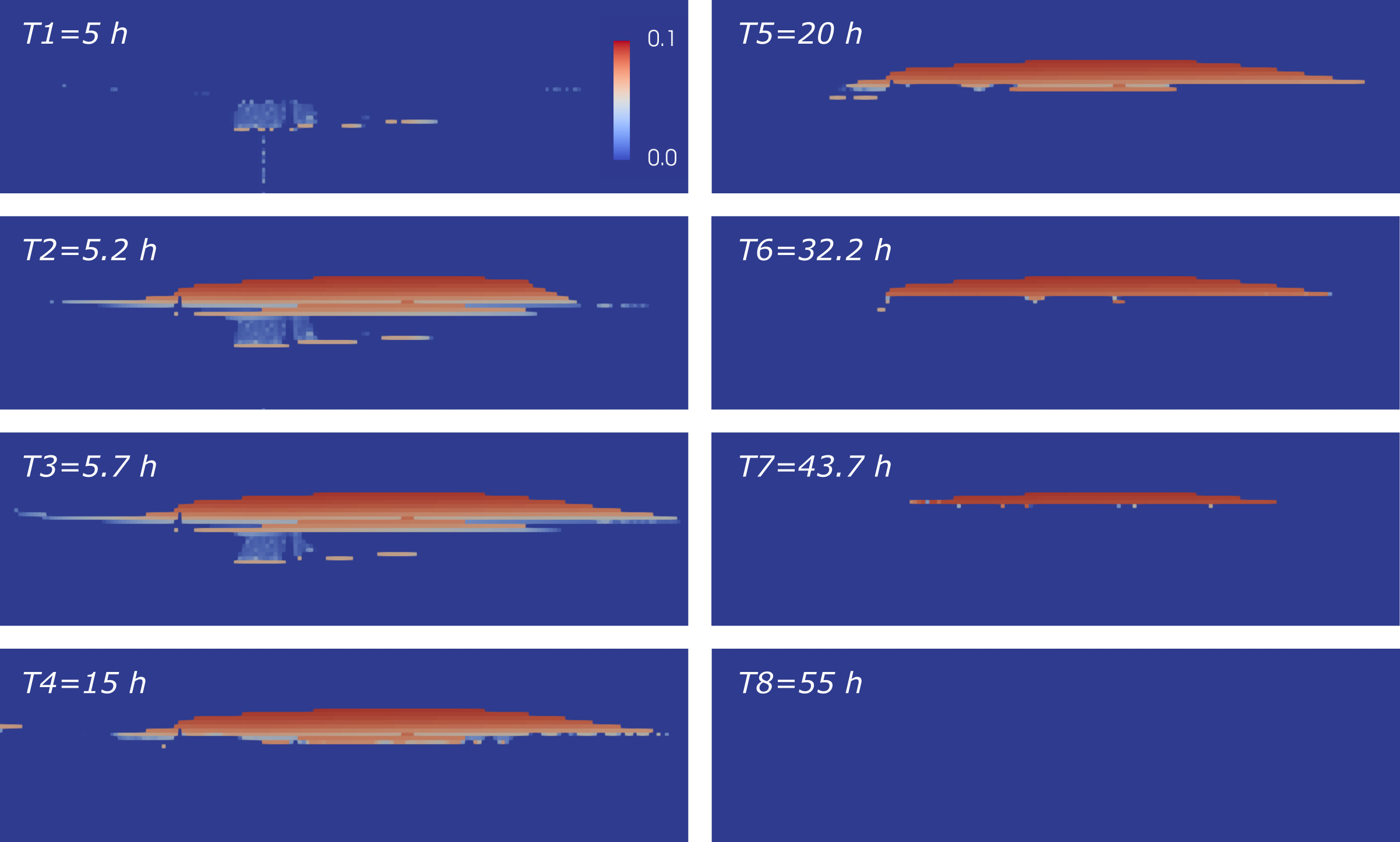}
  \caption{Residual gas saturation profiles in the presence of hysteresis effects.}
\label{fig:Residual gas saturation}
\end{figure}

\autoref{fig:Residual gas saturation} shows the evolution of residual gas saturation in upper gas cap over time. 
During injection period, very few cells have CO2 in residual form because the flow is dominated by viscous force. Immediately after injection stops, the driving force within the system switches from viscous force to a balance between gravity and capillary forces. Consequently, the drainage process occurring in most cells is interrupted and transitioned to imbibition.
In post-injection period, brine can be imbibed into CO2 plume, which favours the trapping of CO2 in residual form. Residual gas is further dissolved into brine: its region is found to shrink and even disappear as the convective transport proceeds.


It should be pointed out that the inclusion of hysteresis effect may seem trivial on plume migration, especially in cases where the primary drainage and imbibition curves look similar; it gives more accurate prediction of residual trapping, which helps to assess when injected CO2 is securely trapped or stored in a more permanent form.

\subsection{Impacts of diffusion}
\begin{figure}[h]
  \centering
  \includegraphics[width=0.9\linewidth]{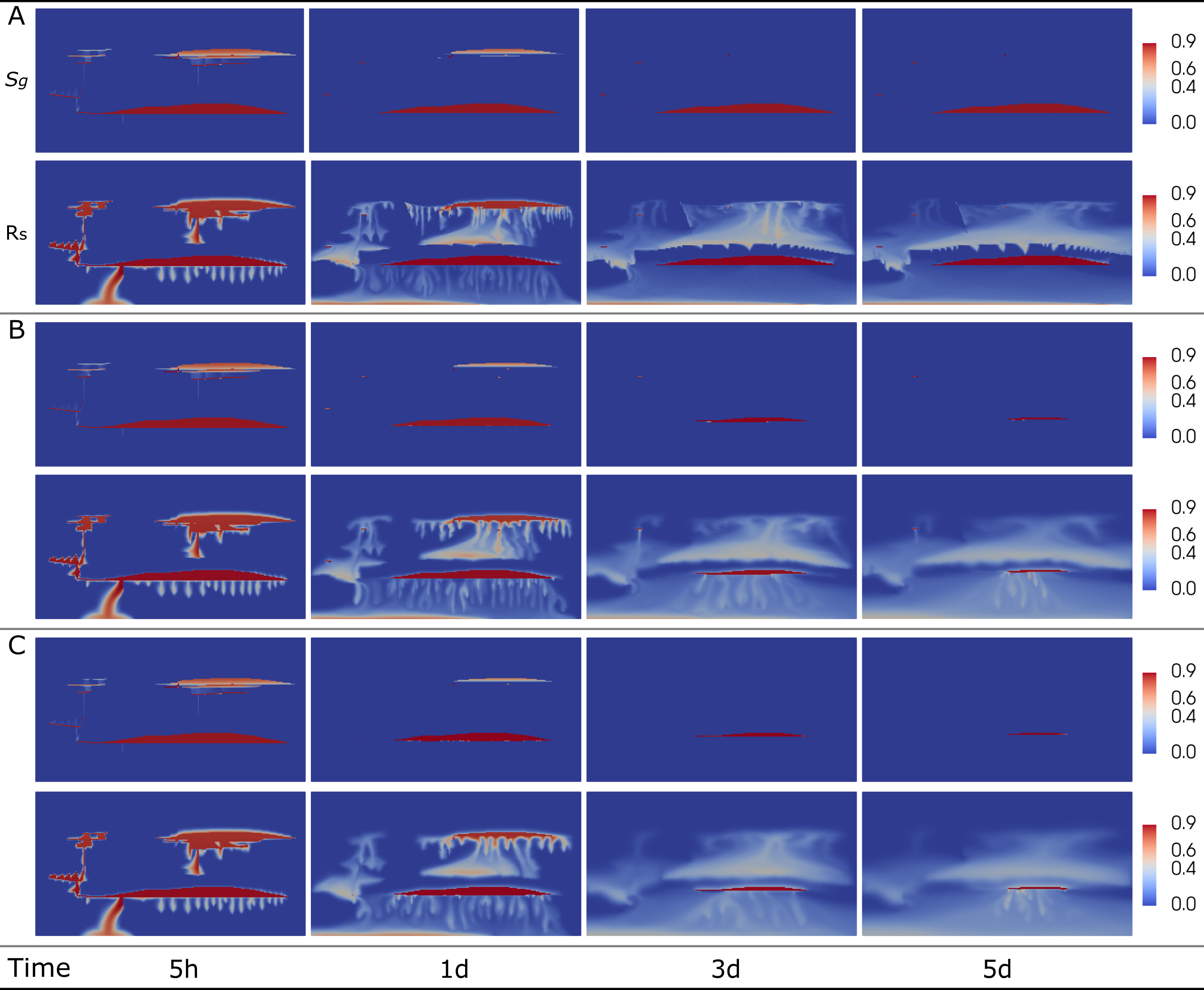}
  \caption{Gas phase saturation and solution CO2–brine ratio profiles in the absence or presence of diffusion effect. (A), (B), and (C) show results for D=0 $m^2/s$ (base case), D = 2e-9 $m^2/s$, and D = 5e-9 $m^2/s$, respectively.}
\label{fig:Diffusion results}
\end{figure}

To investigate the impact of diffusion, the diffusion coefficient of CO2 in liquid phase, i.e., $D_{\text{CO}_2, l}$, is increased from 0 to positive values. 
\autoref{fig:Diffusion results} shows the gas saturation and solution ratio maps for the three investigated cases. 
At the time when injection stops, the two cases considering diffusion effect do not show large difference compared to the base case, because viscous force dominates the displacement process.
During post-injection period, both the lower and upper gas cap observe an enhancement of convective transport in the presence of diffusion, indicated by smaller gas caps at the same time. 
We also observe that interfaces between descending fingers and surrounding brine are more diffusive in cases (B) and (C).
CO2 continues to leave the lower gas cap in the form of dissolved fingers, and the gas cap almost disappears after 5 days.

\begin{figure}[h]
  \centering
  \includegraphics[width=0.9\linewidth]{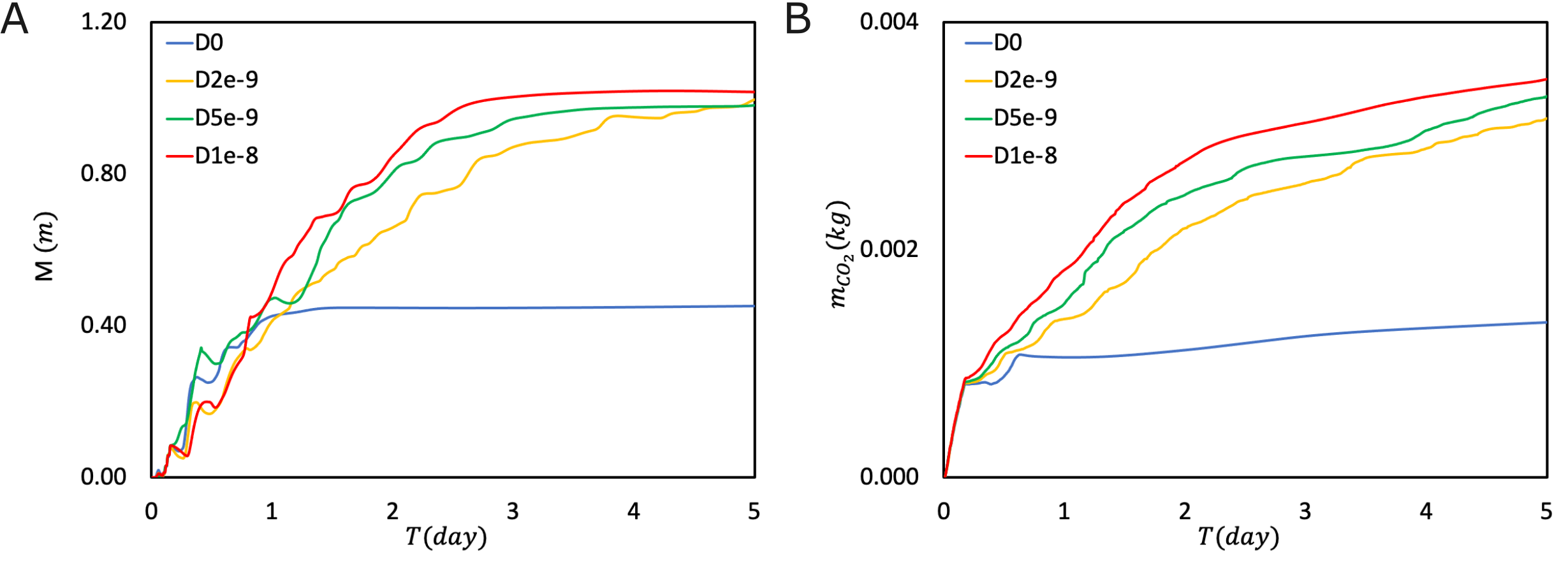}
  \caption{M value (A) and dissolution trapping of box A (B) in the absence or presence of diffusion effect.}
\label{fig:Diffusion_Mc & dissolution trapping}
\end{figure}

\autoref{fig:Diffusion_Mc & dissolution trapping}A presents the dynamic behavior of $M$, which serves as a measure of the degree of convective mixing and is given by:
\begin{equation}
M(t)=\int \left \lvert \nabla \frac{x_{c}^{w}}{x_{c, max}^{w}} \right \rvert dx,
\end{equation}
where $x_{c}^{w}$ and $x_{c, max}^{w}$ denote the mass fraction of CO2 in liquid phase and the dissolution limit, respectively. As shown, inclusion of diffusion effect results in a remarkable increase of $M$. A larger diffusion coefficient leads to a stronger convective mixing under the lower gas cap, yet the stabilized value of $M$ is close regardless of the strength of the diffusion coefficient.
This finding is also consistent with the quantification results of dissolution trapping in box A. As shown in \autoref{fig:Diffusion_Mc & dissolution trapping}B, the largest amount of dissolved CO2 is found in the case of D=1e-8 m$^2$/s, and the three cases with nonzero diffusion coefficient tend to have similar increase rate at late stage.

\section{Conclusion} 
In this work, a unified framework is developed to model the essential trapping physics of CO2 storage in saline aquifers across varying time scales.
Particularly, a parameterization method is proposed to describe hysteretic behavior of constitutive relations via interpolation of the parameterized space, which are constructed based on current saturation and the turning point saturation values. Moreover, analytical expressions are derived for thermodynamic equilibrium calculation using the black-oil fluid model. Partitioning of components into phases are readily obtained without performing standard flash calculations.

The proposed framework is validated by investigating the behavior of gravity-induced convective transport. Results indicate that both the fingering pattern and the quantitative dissolution rate agree well with those reported in literature. The numerical simulator is then applied to study the FluidFlower benchmark model with well-defined geological heterogeneity. Physical parameters, i.e., the permeability of different sand layer, are tuned such that the simulation results match physical observations from the tracer test.  
Results indicate that hysteresis effect does not have a great impact on the migration of plume in this specific setting, and most of the residually trapped CO2 is eventually dissolved into brine.
Furthermore, inclusion of diffusion into the model enhances the convective mixing thereby increasing the amount of dissolution trapping.
Our developed numerical model has shown to capture the relevant physical processes for CO2 storage in saline aquifers, which is promising to be further applied to large-scale problems of interest.

\backmatter





\bmhead{Acknowledgments}

Yuhang Wang was supported by the ``CUG Scholar'' Scientific Research Funds at China University of Geosciences (Wuhan) (Project No.2022157). Ziliang Zhang acknowledges the financial support of China Scholarship Council (No.202207720047). Hadi Hajibeygi was sponsored by the Dutch National Science Foundation (NWO) under Vidi Talent Program Project ``ADMIRE'' (Project No.17509). We thank all ADMIRE members and its user committee for allowing us publish this paper.

\section*{Declarations}

\bmhead{Conflict of Interest} The authors declare that they have no known competing financial interests or personal relationships that could have appeared to influence the work reported in this paper. 

\bmhead{Authors' Contributions} Conceptualization, Y.W., Z.Z., H.H.; methodology, Y.W., Z.Z., H.H.; software, Y.W., Z.Z.; investigation, Y.W., Z.Z.; writing—original draft preparation, Y.W., Z.Z.; writing—review and editing, C.V., H.H.; supervision, C.V., H.H.; project administration, C.V., H.H.; funding acquisition, Y.W., H.H.

\bibliography{sn-bibliography}


\end{document}